\documentclass[letterpaper,prx,aps,twocolumn,showpacs,superscriptaddress,nofootinbib]{revtex4-2} 

\usepackage{bm,color}
\usepackage[T1]{fontenc}
\usepackage[utf8]{inputenc}
\usepackage{latexsym}
\usepackage{graphicx} 
\usepackage{amsmath} 
\usepackage{amsfonts} 
\usepackage{braket}
\usepackage{subfigure}
\usepackage{bbold}
\usepackage{lineno}
\usepackage{cancel}
\usepackage{mathtools, physics}
\usepackage{hyperref}

	\usepackage[normalem]{ulem} 
	\definecolor{mgreen}{RGB}{1,123,0}
    \definecolor{morange}{RGB}{230,120,20}

\def\be{\begin{equation}}
\def\ee{\end{equation}}

\newcommand{\bk}{\mathbf{k}}

\begin{document}

\title{Adiabatic realization of anomalous Floquet topological systems}

\author{Luca Asteria}
\email{asteria.luca.6w@kyoto-u.ac.jp}
\affiliation{Department of Physics, Graduate School of Science, Kyoto University, Kyoto 606-8502, Japan}

\author{Marcel N. Kosch}
\affiliation{Institute for Quantum Physics, University of Hamburg, 22761 Hamburg, Germany}

\author{Henrik P. Zahn}
\affiliation{Institute for Quantum Physics, University of Hamburg, 22761 Hamburg, Germany}

\author{Jonathan Bracker}
\affiliation{Department of Physics, TU Dortmund University, 44227 Dortmund, Germany}

\author{André Eckardt}
\affiliation{Institut für Physik und Astronomie, Technische Universität Berlin, 10623 Berlin, Germany}

\author{Klaus Sengstock}
\affiliation{Institute for Quantum Physics, University of Hamburg, 22761 Hamburg, Germany}
\affiliation{The Hamburg Centre for Ultrafast Imaging, 22761 Hamburg, Germany}

\author{Christof Weitenberg}
\email {christof.weitenberg@tu-dortmund.de}
\affiliation{Department of Physics, TU Dortmund University, 44227 Dortmund, Germany}


\begin{abstract}
Topology has emerged as a central concept for classifying phases of matter. The situation is especially rich in periodically driven systems, where anomalous Floquet topological phases break the usual bulk-boundary correspondence between Chern number and edges modes of two-dimensional systems. These phases were so far realized by periodic modulation of the tunneling elements at frequencies near-resonant with respect to the system's bandwidth, a regime where Floquet heating plays a significant role in interacting systems. Here we show that such anomalous Floquet topological phases can also be realized by means of an adiabatic protocol, where the system is always in the instantaneous ground state of the cyclic path in parameter space, like in a Thouless charge pump. We experimentally realize such a state using ultracold atoms in a hexagonal lattice where we adiabatically modulate the lattice geometry, including the sublattice offset. To infer the topology, we use the micromotion area in real-space, which was recently identified as a proxy for the winding number. This way of realizing anomalous phases avoids resonant Floquet heating and imperfect loading into the target state. We demonstrate the robustness of the adiabatic construction by observing the anomalous phase even in the presence of mean-field interactions of magnitude comparable to all other energy scales. These findings are promising for engineering novel topological states in a more robust way.
\end{abstract}

\maketitle

Topological arguments explain the quantization of the quantum Hall anomalous conductivity \cite{vonKlitzing1980}, which is at the basis of precision measurements and practical applications \cite{vonKlitzing1986}. It manifests as topological indices, which are usually integer and enforce quantized responses as well as protected states at the edge of the system. Of particular interest is the possibility of controlling the topological properties of a system by means of time-periodic driving (so-called Floquet engineering \cite{Bukov2015, Eckardt2017,Oka2019,Weitenberg2021}). This strategy permits to realize effective static Hamiltonians with interesting novel properties. But in the regime when the driving frequency approaches the energy scale of the unmodulated system, it also allows to engineer band structures and states without static counterpart.
In this latter case, a different topological classification is needed, as the quasienergies $\epsilon$ of Floquet states are defined only up to an integer multiple of $\hbar\omega$ (with $\hbar$ being the reduced Planck constant and $\omega$ the driving frequency), so that they are defined on a ring, giving rise to $n$ band gaps between $n$ quasienergy bands (rather than the $n-1$ gaps between $n$ energy bands of undriven systems), each of which can host chiral edge modes. As a result  Chern numbers, which count the difference of the number edge modes in the gaps above and below the bulk band they are associated with, can vanish in a system with edge modes~\cite{Kitagawa2010,Rudner2013,GomezLeon2024anomalousfloquet,gavensky2025PRX}. The topology is then captured by the winding numbers defined for each band gap, e.g., $W_0$ and $W_\pi$ for a two-band model with gaps centered around $\epsilon T/\hbar=0$ and $\pi$ (with driving period $T$). Such systems have been realized in photonics \cite{Maczewsky2017,Mukherjee2017}, acoustic systems \cite{Peng2016} and cold atoms \cite{Wintersperger2020} and have been detected through propagation of chiral edge modes \cite{Peng2016,Maczewsky2017,Braun2024} or via a combination of spectroscopic and state-tomography methods~\cite{Uenal2019,Wintersperger2020}.  
Moreover, anomalous Floquet topological systems have recently attracted attention because their edge modes do not rely on the existence of delocalized bulk states and, thus, are more robust against disorder compared to those of Chern insulators~\cite{Titum2016,Nathan2019,Hesse2025}. 
Also the loading into edge modes with a quench of the modulation appears easier as compared to Chern systems \cite{Martinez2023,Braun2024,Hesse2025}.  
However, the use of these anomalous modes for practical applications is still limited by technical factors, among which Floquet heating likely provides the greatest contribution. Furthermore, while all measurements mentioned so far were conducted in the absence of significant interactions, Floquet heating effects become more severe when interactions are included \cite{Weinberg2015,Reitter2017,Eckardt2017,Weitenberg2021}.

A Thouless charge pump, as realized for instance in the slowly driven rice Rice-Mele model, is a second example of a driven system where time plays a crucially important role \cite{Rice1982,thouless1983}. 
Here the quantized charge pumping is characterized by a Chern number in the 2D parameter space spanned by time and quasimomentum, which can be computed directly in the adiabatic limit by the number of times the path in parameter space winds around a singularity. It has been measured, among others, also in cold atom systems \cite{Wang2013,Nakajima2016,Lohse2016,Lu2016PRL,Walter2023,Zhu2024_reversal}. 
Robustness (e.g. against quasiperiodic disorder \cite{Nakajima2021,Hayward2021}) is ensured by adiabaticity and does not depend on the exact path in parameter space. Due to the ensuing high fidelity, Thouless pumps can be used for spin-resolved imaging in a quantum gas microscope \cite{Koepsell2020}, might be used to manipulate qubits in a quantum register in an optical lattice \cite{zhu2024quantumcircuitsbasedtopological,kiefer2025}, and also appear in fractional form for solitons \cite{Bohm2026}.

\begin{figure}[h!]
	\includegraphics[width=0.9\linewidth]{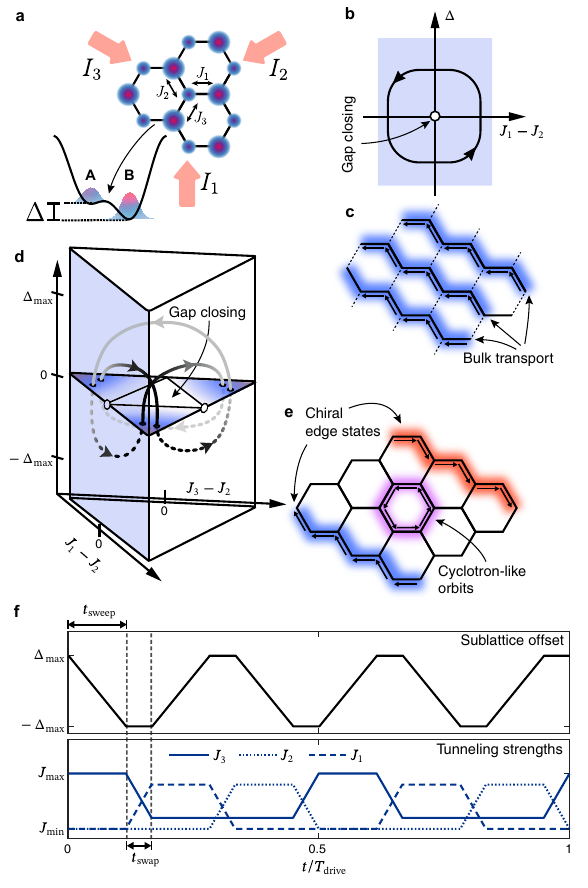}
\caption{ {\bf A 2D charge pump as an anomalous Floquet topological phase.} \small a) We realize a honeycomb lattice with tunnel couplings $J_{1},J_2,J_3$ along the three different directions between neighboring sites, respectively, formed by three laser beams with intensities $I_{1},I_2,I_3$ and a sublattice offset $\Delta$ (represented by disks of different radius). b) For $J_3=0$ the system corresponds to uncoupled Rice-Mele-type 1D chains, described by a parameter space $(J_1-J_2,\Delta)$ with a band degeneracy at $(0,0)$. c) A cyclic path around the degeneracy realizes the well-known quantized Thouless pump along these 1D chains.
d) When including $J_3$, the parameter space becomes three dimensional (including $J_1-J_2$ and $J_2-J_3$ with the constraint $\sum J_i=const.$) and has band degeneracies (associated with Dirac points) within the central triangular area in the $\Delta=0$ plane (small white triangle). Now new nontrivial paths that cannot be continuously deformed to a single point become possible. The simplest nontrivial path involves three segments with $\Delta>0$ (solid lines) and three segments with $\Delta<0$ (dashed lines). e) The corresponding driving protocol is an adiabatic version of the anomalous Floquet topological system, and contains circular trajectories in the bulk as well as chiral edge states. f) The experimental protocol has six steps, each of which consists of one offset sweep and one swap of the strongest tunnel element.}    \label{fig:aAFTI_sketch}
\end{figure}

\section{A 2D charge pump as an anomalous Floquet topological phase} 
Here we draw a new connection between charge pumps and anomalous Floquet topological phases. We do so specifically in a tunable honeycomb lattice with tunnel couplings $J_1$, $J_2$ and $J_3$ along the three directions connecting neighboring sites and energy offset $\Delta$ between the A and B sublattice (Fig.~\ref{fig:aAFTI_sketch}a). To understand the underlying concept, let us first recapitulate usual Thouless pumping in the context of such a lattice. When setting one tunnel coupling to zero, say $J_3=0$, we obtain many decoupled 1D chains, described by the parameter space $(J_1-J_2,\Delta)$. An adiabatic cyclic path in this parameter space around the band closing point at $(0,0)$ then describes the well-known Thouless charge pump (Fig.~\ref{fig:aAFTI_sketch}b,c). In the honeycomb lattice, three equivalent embeddings of a 1D charge pump are possible by setting one of the tunnel elements to zero and modulating the other two. We note that also for finite coupling between the chains $J_3\ne0$ (e.g. $J_3=0.15 J_{\rm max}$ in the experiments below), the corresponding path remains topologically non-trivial, and still produces quantized transport along the 1D chains, as long as the coupling does not close the band gap (Supplementary Information~\ref{supp:Chern_T}).

Leaving the regime with one tunnel parameter reduced relative to the others, we have a 3D parameter space $(J_1-J_2,J_3-J_2,\Delta)$ when imposing the constraint $\sum J_i=\text{const}$ (Fig.~\ref{fig:aAFTI_sketch}d). In the $\Delta=0$ plane, inversion symmetry enforces the bands to touch at the Dirac points, which are present in the central triangular area of gap closing in Fig.~\ref{fig:aAFTI_sketch}d. A band gap can only open via the merging transition of the two Dirac points~\cite{Zhu2007,Tarnowski2017}, which happens at the edges of this triangular area, where one tunnel coupling is equal to the sum of the other two. Furthermore, at the vertices of the $(J_1-J_2,J_3-J_2,\Delta=0)$ parameter space, one tunnel coupling completely dominates over the other two and the system consists of a 2D array of decoupled double wells. 
In this parameter space, new nontrivial paths that cannot be continuously deformed to a single point become possible. Of particular interest is the path sketched in Fig.~\ref{fig:aAFTI_sketch}d. It does not generate a net center of mass transport, as every allowed region in the $\Delta=0$ plane is traversed twice in opposite direction. But, as the central result of this article, we find that the drive corresponding to this path generates an anomalous topological phase. It is topologically equivalent to the anomalous Floquet topological phases as they are conventionally realized via a near-resonant modulation of the tunnel couplings \cite{Rudner2013,Kitagawa2010,Wintersperger2020}, except that the offset sweeps double the period of the protocol to six steps and accordingly the winding numbers to 2 (see also Supplementary Information~\ref{adiabatic_Winding_Number} for details). This protocol, thus, constitutes an \emph{adiabatic} realization of an anomalous Floquet topological phase. Just like in a 1D Thouless pump, such a path breaks time-reversal symmetry by the direction along which the path is followed.

The adiabatic paths have a simple interpretation in real space. Assuming tunneling to be suppressed in two directions, during an adiabatic sweep of the offset $\Delta$ from a large positive to a large negative value, a particle originally localized in the lower B site stays in the instantaneous ground state and moves to the neighboring A site along the strong tunnel bond. The repetition of such sweeps with different tunnel matrix elements $J_i$ being enhanced allows to adiabatically transport the particles along desired paths. Swapping between strong $J_1$ and strong $J_2$, we achieve 1D Thouless-type charge pumping in one direction. In turn, the second protocol shown in Fig.~\ref{fig:aAFTI_sketch}d is obtained when cycling through the three tunnel directions, e.g., by making $J_3$, $J_1$, $J_2$, $J_3$, $J_1$, $J_2$,\ldots strong. As a result, the lattice returns to its initial configuration after six steps, since the sublattice offset $\Delta$ is reversed after three steps (see Fig.~\ref{fig:aAFTI_sketch}f). In the bulk, this leads to a circular motion around a hexagonal plaquette of the lattice within one driving cycle, bringing the particle back to its initial state, corresponding to a dispersionless Floquet-Bloch band.  At the edge, however, every third step no transport occurs, inducing directed chiral transport along the edge akin to a 1D Thouless pump (Fig.~\ref{fig:aAFTI_sketch}e).

We will show in this article that the adiabatic realization of the anomalous topological phases is significantly more robust than the conventional realization via near-resonant tunnel modulation.  The robustness concerns (interaction-induced) Floquet heating, interactions and disorder (the last of which we study numerically at the end of the article) {and holds even if imperfect adiabatic transport could happen at very long timescales \cite{Privitera2018}}. An important Floquet heating mechanism are resonant excitations of collective modes in interacting systems \cite{Eckardt2017}. While the near-resonant scheme has a Floquet driving frequency on the scale of the tunneling energy $J$, the adiabatic realization can be implemented with arbitrarily small frequencies, where these resonant excitations play almost no role. This explains the observation that our experimental realization works despite interactions of magnitude comparable to all other energy scales, while implementations of the near-resonant scheme required Feshbach resonances to tune the scattering length close to zero \cite{Braun2024}.

Moreover, even though the driven system is characterized by a non-trivial topology, an efficient loading of the bulk states is possible directly from the trivial unmodulated system, as both systems are characterized by the same eigenstates, without the need of crossing band-gap closing points as they occur when tuning an initially static system into a Chern or anomalous Flouquet topological system \cite{budich2015,DAlessio2015,barbarino2020,Wintersperger2020}.

\begin{figure}[h!]
	\includegraphics[width=0.9\linewidth]{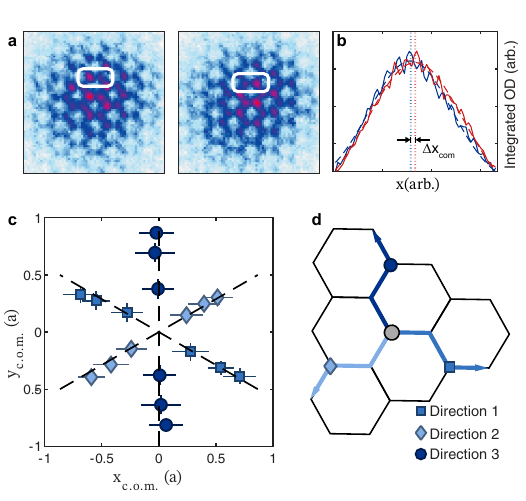}
	\caption{{\bf 1D charge pumps embedded in a 2D lattice.} 
    a) Images of the atoms at the beginning and end of a driving cycle as in Fig.~\ref{fig:aAFTI_sketch}b obtained with the matter-wave microscope after ramping to a balanced situation $I_1=I_2=I_3$. The populations of the two sublattices periodically invert after each step as becomes visible in the white rectangle containing an A and a B site. b) Overlay of the cut through the two images illustrating the displacement of the center of mass $\Delta x_{\rm CoM}$. The parameters are $\Delta=2\pi\cdot7\,\mathrm{kHz},~\varepsilon=0.5,~I_\mathrm{max}=7~E_r,~t_{\rm sweep}=1~\mathrm{ms},~t_{\rm swap}=0.45~\mathrm{ms}$. c) Movement of the center of mass for 1, 2, and 3 pump cycles for three different protocols, leading to a transport along three directions (dashed lines) as illustrated in d). Here $t_{\rm swap}=1~\mathrm{ms}$. The errorbars show the standard deviation from the mean with about $\sim7$ measurements.} 
    \label{fig:2D_Thouless}
\end{figure}

\section{Experimental realization in a hexagonal optical lattice}
We experimentally realize the adiabatic protocols introduced above using a Bose-Einstein condensate of $^{87}$Rb atoms loaded into a hexagonal optical lattice formed by the interference of three laser beams (Fig.~\ref{fig:aAFTI_sketch}a). No lattice is applied in transverse direction, so that we realize a 2D lattice of 1D tubes.

Varying the intensities $I_i$ of the three laser beams, we can tune the tunnel parameters $J_i$ (see Methods~\ref{optical_lattice}). We denote the maximum and minimum beam intensities $I_\mathrm{max}$ and $I_\mathrm{min}$, respectively (via their resulting optical potential in units of recoil energy) and define the maximum beam imbalance $\varepsilon$, so that ${I_{\min}=}(1-\varepsilon) I_\mathrm{max}$. To reduce the parameter space, we also impose $I_\mathrm{tot}=\sum_i I_i=const.$ One dominant intensity, say $I_3$, makes the tunnel coupling $J_3$ dominate over $J_1$ and $J_2$. Furthermore, we dynamically control the sublattice offset $\Delta$ with the multi-frequency optical lattice scheme introduced in \cite{Kosch2022}. This allows us to implement both of the adiabatic protocols sketched above.

We initialize the system by loading the bosonic atoms into the lattice at strong sublattice offset $\Delta=\Delta_\text{max}\gg J_i$. In this regime, the A and B sublattices roughly correspond to the upper and lower band, respectively, and the bands become approximately flat, as tunneling within each sublattice can be neglected. In the initial state the atoms predominantly occupy the B sublattice, i.e., the lowest band. Importantly, moreover, the coherences between different sites decay via dephasing, thanks to the lack of inter-sublattice tunneling. Both effects together lead to an initial state that resembles that of a fermionic band insulator, with the particles being delocalized with respect to quasimomentum within a single band, while also being delocalized in real space. Such a state can also be viewed as a product state in real space, which provides the intuition for the adiabatic protocols sketched above.

\section{1D charge pumps embedded in a 2D lattice} 
We demonstrate the control over our system by realizing the 1D Thouless pumps embedded in a tunable honeycomb lattice as discussed above. To detect the directed transport, we measure the center-of-mass position of the cloud with very high resolution using the matter-wave microscope \cite{Asteria2021} with a magnification of $82$, which yields an effective resolution below $100$\,nm and allows to resolve the lattice structure (Fig.~\ref{fig:2D_Thouless}a). We fit the position in the $x$ ($y$) direction with a Gaussian function after integrating the signal in $y$ ($x$) direction, and obtain the position of the center of mass (Fig.~\ref{fig:2D_Thouless}b). 
We check that by reversing the sign of $\Delta(t)$, the cloud moves in the opposite direction (Fig.~\ref{fig:2D_Thouless}c).
We also observe a $120^{\circ}$-rotation of the transport direction after cyclic permutation of the beam indices. 
The errors bars are limited by the shot-to-shot fluctuations of the position of the magnetic trap center with respect to the imaging camera of order $\sim6~\mu$m (corresponding to $\sim70~$nm in the magnified image) and just allow to resolve the transport. 

In our experiment, we observe a displacement which is a factor 2.7 smaller than expected for the quantized value. We attribute this deviation mainly to a finite occupation of the second band at $t=0$, which is transported in the opposite direction. We elaborate in the Methods~\ref{breakdown_effects} on several additional effects that might contribute.

\begin{figure}[htb]
	\includegraphics[width=0.95\linewidth]{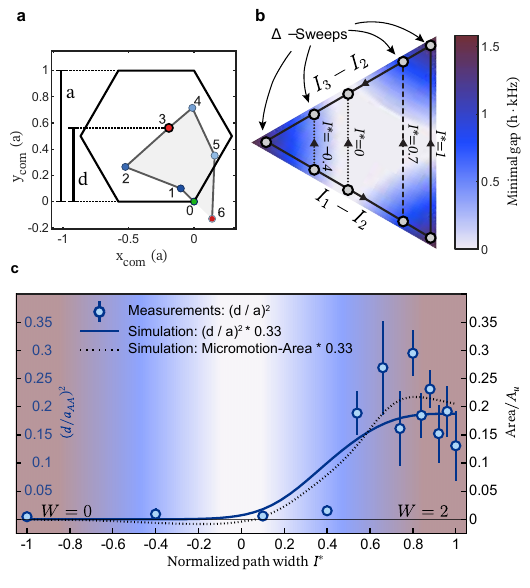} 
	\caption{{\bf Adiabatic realization and detection of anomalous Floquet topology.} a) Measured micromotion trajectory for the adiabatic realization of the anomalous Floquet topological system (with $\varepsilon=0.45$). The hexagonal pattern indicates the unit cell. $d$ is the projection along $y$ of the position of the center of mass after three steps (relative to $t=0$).
    b) Projection of the path of the driving protocol into the $\Delta=0$ plane. The minimal instantaneous band gap {during the protocol} is indicated by the color code. In each protocol, we start at the left corner and move along the equilateral triangle of side $(I_1-I_2)^\ast$. At each corner, we perform a sweep inverting the offset with $\Delta_{\max}=h\cdot2.6$\,kHz. The other parameters are $t_{\rm sweep}=1~\mathrm{ms},~t_{\rm swap}=0.45~\mathrm{ms}$.
    c) Measured micromotion area $(d/a)^2$ as a function of the path width $(I_1-I_2)^\ast$ (blue data points). Errors bars represent the standard deviation on the mean of 6 individual measurements. We plot the corresponding interaction-free theory (scaled down by a factor 0.33) evaluated as full area (dashed line) and as $(d/a)^2$ (solid line). The colormap indicates the size of the smallest instantaneous band, at which the protocol crosses the $\Delta=0$ plane.}
    \label{fig:triv_2_AFTI}
\end{figure}

\section{Adiabatic realization of anomalous Floquet topology}
\label{experiment_adafti}
We now turn to the realization of an anomalous Floquet topological phase using the same experimental setup via the protocol introduced in Fig.~\ref{fig:aAFTI_sketch}d. We characterize the anomalous Floquet topological phase via the micromotion area traced out by localized particles \cite{asteria2026micromotionareaproxyanomalous}. For a lattice with two-sublattice states, the anomalous Floquet topological phase with equal winding numbers $W_0=W_\pi$ is characterized by the area $A$ traced by the center of mass of an initially localized particle during one driving period: for dispersionless dynamics, where Floquet-Bloch bands are completely flat (dispersionless point), the area is quantized in terms of the winding numbers, $2A/A_u=W_0=W_\pi$, where $A_u$ is the unit cell area \cite{asteria2026micromotionareaproxyanomalous}. However, also away from this point, a micromotion area $A$ on the order of one in units $A_u$ still serves as a proxy for anomalous topology \cite{asteria2026micromotionareaproxyanomalous}. In contrast, the micromotion area tends to values much smaller than one away from the anomalous Floquet topological phase. It is therefore a unique indicator of the anomalous Floquet topological phase compared to both topologically trivial phases and Chern phases.

As described above, we start with many particles occupying several lattice sites, corresponding to a sample with the particles initially completely localized on the B sites. Therefore the dynamics for all particles starting from the same sublattice is exactly the same and we can access it by measuring the dynamics of the center of mass of the cloud. To evaluate the area from noisy experimental data, we measure $d$, the projected distance along the $y$ direction traveled by the center of mass after three steps (i.e.\;half a full driving cycle) and approximate $A\approx(d/a)^2$ with the lattice constant $a$, i.e., the distance between neighboring unit cells. This allows us to restrict the data taking to only two time steps $t=0$ and $t=T_{\rm drive}/2$ for further parameter scans below. For the ideal hexagonal path, $(d/a)^2$ is identical to the micromotion area $A$, and our numerics shows that it remains a very good approximation for the occurring paths with smaller areas (compare solid and dashed lines in Fig.~\ref{fig:triv_2_AFTI}c). To limit systematic effects, $d$ is obtained as a differential quantity by measuring half of the points for $\Delta(t)\rightarrow -\Delta(t)$, which sends $d\rightarrow-d$. 

We find indeed a finite micromotion area of about $A=0.3 A_u$ (Fig.~\ref{fig:triv_2_AFTI}a). This value is significantly reduced compared to $A=W_0(A_u/2)=A_u$ corresponding to the winding numbers $W_0=W_\pi=2$, but nevertheless it clearly allows us to identify the anomalous Floquet topological regime, as it drops to vanishingly small values in the topologically trivial regime (see below). We attribute this reduction to a number of effects including interactions and finite adiabaticity (compare Methods~\ref{breakdown_effects} and \ref{Kitagawa_A_W}). The signature is robust against the exact shape of the path and we verify that we measure a finite micromotion area for different values of the maximal beam imbalance $\varepsilon$ as long as the path does not cross the gap closing (Extended Data Fig.~\ref{fig:Area_vs_I}).

After having identified the anomalous regime, we investigate the topological transition that occurs when deforming the cyclic path to a trivial protocol (Fig.~\ref{fig:triv_2_AFTI}b). We keep the starting point in the left corner fixed and vary the side of the equilateral triangle of the path projected on the $\Delta=0$ plane, which we label as path width $I^\ast :=  (I_1-I_2)^\ast /(\varepsilon I_{\max})$. In the parameter space $(I_1-I_2,I_2-I_3,I_3-I_1)$, the tunnel couplings do not reach zero and the band gap closes in a finite region for $I_1-I_2$ and $I_2-I_3$. The cyclic path therefore crosses the $\Delta=0$ plane at closed bands for a range of path widths $-0.09<I^\ast<0.09$, in which the path cannot be adiabatic. To the left of this transition region, the path is trivial as it can be contracted to a single point and we expect a topologically trivial Floquet system. 

We characterize this topological phase transition via the change of the winding numbers obtained from the micromotion dynamics. We find indeed that for paths with path widths below the transition region, $(d/a)^2$ drops to zero, indicating the trivial topology (Fig.~\ref{fig:triv_2_AFTI}c). We compare the curve to the micromotion area and $(d/a)^2$ from an interaction-free theory of the full time-evolution of the micromotion dynamics, which are scaled down by a factor 0.33 to account for systematic effects (see Methods~\ref{breakdown_effects}). Agreement between data points and numerical curves confirm the observation of the adiabatic anomalous Floquet topological phase via the micromotion area.

\begin{figure}
    \includegraphics[width=0.95\linewidth]{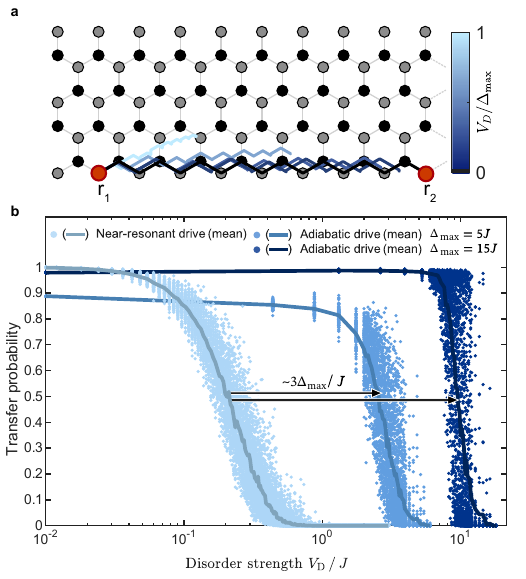}
    \caption{{\bf Numerical study of robustness to disorder}. 
    a) Geometry of the numerical simulation for the transport along a zig-zag edge of a honeycomb lattice. The particle is initially located at site $r_1$ and we study the probability $|G(r_1,r_2)|^2$ of arriving at site $r_2$ after exactly $N_{\mathrm{steps}}=16$. This corresponds to 8 Floquet periods for the near-resonant drive and 4 Floquet periods for the adiabatic drive. The curves show the center-of-mass trajectories of a particle initially localized on a single lattice site for different disorder strengths and $\Delta_{\max}=15 J$.
    b) Transmission probability $|G(r_1,r_2)|^2$ as a function of the disorder strength for 100 realizations of the disorder together with an average curve (dark blue points and curve), plotted for $\Delta_{\mathrm{max}}=5J$ and $\Delta_{\mathrm{max}}=15J$.
    The same quantities are calculated for the near-resonant drive close to the dispersionless point (light blue points and curve). Here the transfer decays to $0.5$ when the disorder strength approaches $J/5$.}
    \label{fig:AFTI_RABI_disorder_comparison}
\end{figure}

\section{Robustness to disorder}
\label{disorder}
\label{supp:disorder}
Finally we discuss the robustness of the adiabatic protocol to disorder. As we do not have disorder implemented in the experiment, we show results of numerical simulations. We study the dynamics at an edge and record the transmission probability from one initial site, $r_1$, to a target site, $r_2$, both located at the system's edge \cite{Titum2016}. We introduce local disorder in the form of random on-site potentials $V_r$ on sites $r$ sampled uniformly in the interval $[-V_D,V_D]$ and measure the average transmission probability $|G(r_2,r_1)|^2=|\bra{r_2}U(t_{12})\ket{r_1}|^2$ where $U(t)$ is the time-evolution operator. The time $t_{12}$ is chosen such that $|G(r_2,r_1)|^2\sim1$ for the disorder-free case $V_D=0$. We plot the values of $|G(r_2,r_1)|^2$ for 100 disorder realizations together with an average curve (Fig.~\ref{fig:AFTI_RABI_disorder_comparison}). 
The plotted results include simulations, where the anomalous phase is realized using the standard near-resonant Floquet drive. Here, the transmission drops to $0.5$ as soon as $V_D$ becomes comparable to $\sim0.2J$. In contrast, for our adiabatic protocol, the transmission doesn't drop to $0.5$ until $V_D\sim0.6\Delta_{\max}$ for $\Delta_{\rm max}/J>10$. Because $\Delta_{\max}$ can be tuned to be arbitrarily bigger than $J$, this suggests that the adiabatic realization is significantly more stable against disorder. 

Similar enhanced robustness against disorder using Landau-Zener sweeps was also numerically found for state transfer in one-dimensional systems \cite{Longhi2019AQT,Longhi2019PRB}. Note that while away from the dispersionless point a weak disorder can potentially stabilize the anomalous topology \cite{Titum2016,Hesse2025}, we find no similar effect for both types of drives in or close to the dispersionless point. We suspect that the dispersionless point already provides the most robust protocols.

\section{Discussion}

We have proposed a new way of realizing anomalous Floquet topological band structures in a two dimensional honeycomb lattice. It is based on adiabatic transport, similar to the dynamics realized in Thouless pumps in one dimension. Moreover, we implemented and tested the approach experimentally, using ultracold Rubidium atoms confined in a two-dimensional optical honeycomb lattice. For that purpose we estimated the area enclosed by the center of mass motion during each driving cycle, which was recently proposed as a proxy for the winding number characterizing the anomalous Floquet topological phase. Furthermore, we have provided numerical evidence that the adiabatic anomalous Floquet topological insulator is more robust against uncorrelated disorder than previous realizations based on near-resonant driving. Our findings are of direct relevance for the design of robust waveguides.

Thouless pumps in 2D can be used to simulate 4D quantum Hall physics \cite{Zhang2001_4DQHE,Lohse2018} and higher-order topological systems \cite{Benalcazar2017,petrides2020,Wienand2022}, and it would be interesting to study these effects also in the context of this non-separable geometry. In particular, protocols to realize higher-order topological states based on near-resonant drive \cite{Huang2021} could be made more robust by using the corresponding adiabatic drive. 
The protocol could be made more robust by short-cuts to adiabaticity \cite{Torrontegui2013} such as varying sweep rates adapted to the instantaneous energy gap \cite{Andrade2021} or counterdiabatic driving.

Also, charge pumps in a 2D system might allow the realization of a denser circuit for quantum computation as compared to 1D systems \cite{zhu2024quantumcircuitsbasedtopological,kiefer2025}. 
We have demonstrated robustness against interactions of magnitude exceeding the tunneling energy, where for the implementation of anomalous Floquet topological phases based on near resonant driving significant Floquet heating is expected and where previous experiments where performed at zero or small interactions \cite{Maczewsky2017,Mukherjee2017,Wintersperger2020,Braun2024}. This is promising for further studies involving strongly correlated systems. 
In our experiment, mean-field interactions strongly modify the system description, but the underlying anomalous Floquet topological properties are still retained. 
It would be interesting to study whether interactions induce novel topological effects, such as topological solitons \cite{Mukherjee2020} also in the case of an adiabatic drive. 

{\it Data availability - }
Experimental data and analysis code are available at \cite{ReplicationData}.

{\it Acknowledgments - }
The authors acknowledge support by the Deutsche Forschungsgemeinschaft (DFG, German Research Foundation) via the Research Unit FOR 5688 (Project No. 521530974) and via the cluster of excellence AIM, EXC 2056 (Project No. 390715994). K. S. acknowledges funding via 'Hamburg Quantum Computing', financed by EU within EFRE and FHH.



%

\section{Methods}

\setcounter{figure}{0}
\renewcommand{\figurename}{Extended Data Fig.}

\subsection{Micromotion area and winding number as a function of driving period}
\label{Kitagawa_A_W}
\begin{figure*}[htb]
\includegraphics[width=\linewidth]{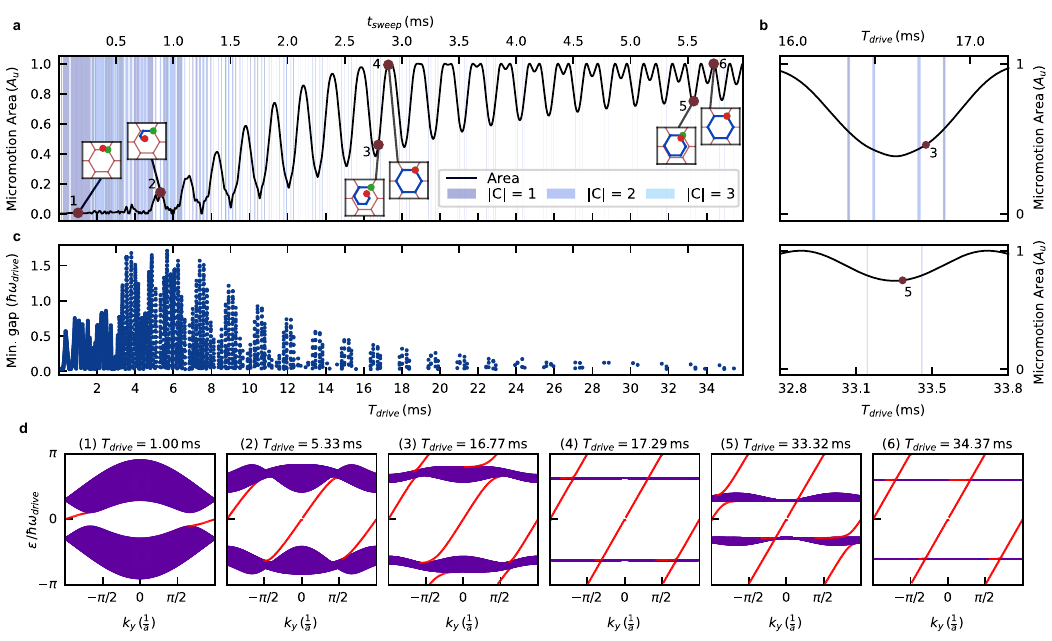}
\caption{{\bf Micromotion area and edge modes as a function of driving period}. 
a) Micromotion area as a function of driving period $T_{\rm drive}=6 T_{\rm sweep}$ for a protocol with stepwise swapping of the tunneling strengths and for $J=2\pi~1.0$\,kHz, $J_\mathrm{min}=0$, $\Delta_{\max}=15J$, and $t_\mathrm{swap}=0$. Simulations are performed for 20,000 values of $T_{\rm drive}$ between 0 and 36\,ms. The area increases towards $A_u$ for large driving periods. When the area begins to rise (above $T_{\rm drive}\gtrsim 6$\,ms), the system is mostly in the anomalous Floquet topological phase with $W_0=W_\pi$ and zero Chern number except for narrow regions around 'commensurate' driving periods. Intervals with non-zero Chern numbers are indicated by blue shading, and are present also in the far right region of the plot, but are hardly visible due to their very small width. Insets show the centre-of-mass trajectories for 6 chosen sample driving periods. b) Zoom into the area-curve in the region around sample points (3) and (5). c) $\pi$-band-gaps in the regions with finite Chern numbers. The band gaps decrease for larger driving periods. These regions come in close-by pairs centered around the local minima of the area curve in a). d) Spectra of the quasienergy as function of quasimomentum $k_y$ calculated for a cylinder geometry with finite-size (160 sites) in $x$-direction and zigzag edges in the periodic $y$-direction. The driving period is indicated above the plots and marked with red markers in a). The color of the points is determined by the average position of the eigenvector along the finite dimension, with red being localized on one edge. We exclude the quasienergies of states mainly localized on the opposite edge. 
\label{PBC_area}}
\end{figure*}

We numerically study the system as a function of the driving period $T_{\rm drive}$. We consider a system with periodic boundary conditions in one dimension (cylinder geometry), which allows to visualize the edge modes and directly put into connection the winding number with the enclosed area. In Extended Data Fig.~\ref{PBC_area}, we plot the resulting behavior of the micromotion area as well as exemplary spectra indicating the bulk bands and edge modes. We call the area encircled during one driving period the micromotion area, although it is also determined by the effective Hamiltonian and its quasienergies, and not by the micromotion operator alone, in order to stress that it requires observing the dynamics within one Floquet period. 
We observe a general trend of the area increasing for longer drive periods, where the driving becomes more adiabatic and gets closer to dispersionless micromotion dynamics, at which the micromotion area is quantized in terms of the winding numbers $W_0=W_\pi$, as already established for the case of a near-resonant drive~\cite{asteria2026micromotionareaproxyanomalous}. For the adiabatic drive, we have $W_0=W_\pi=2$ and the micromotion area is therefore 1 in units of the unit cell $A_u$.

The curve of the micromotion area as a function of the driving period oscillates with a period of around 1.5 ms, which corresponds to oscillations in the transfer efficiency in a sweep starting at a finite value of $\Delta$ (Extended Data Fig.~\ref{PBC_area}). 
The quasienergy distance of the two bands is mainly given by the relative phase evolution in the two sublattices, i.e., on average half the maximum offset $\Delta_{\rm max}/2$. The quasienergy distance is correspondingly $\Delta_\mathrm{max}/2~\mathrm{mod}~\hbar\omega_\mathrm{drive}$ and therefore rapidly changes as a function of the driving period, six times as fast as the oscillation of the area curve. We observe that the dispersion of the edge states is mostly unaffected by the constantly shifting position of the bulk bands, which deform it only slightly. This is in particular true, when the bands are flat which coincides with the large values of the micromotion area (Extended Data Fig.~\ref{PBC_area}). Recall that in the case of a modulated square lattice with near-resonant drive, the two bands exactly fall on top of each other and create an anomalous Floquet topological phase \cite{Rudner2013}. 

We find special 'commensurate' driving periods at which the band gap opens at the $\pi$ gap without edge states, corresponding to a Haldane phase with $W_0=2,~W_\pi=0$. This is not experimentally relevant as it occurs only for very narrow windows of driving periods {(see insets in Extended Data Fig.~\ref{PBC_area}b)}. Furthermore, the band gap in these narrow regions decreases with longer driving periods and reaches small values around $0.1$\,rad for $T_{\rm drive}>30$\,ms (see Extended Data Fig.~\ref{PBC_area}b). 

The oscillation does not prevent us from clearly identifying the anomalous topology from the micromotion area and we therefore find that it is a suitable proxy for anomalous Floquet topology for sufficiently slow, i.e., adiabatic drives. It does not completely damp for longer driving periods $T_{\rm drive}$, but it can be reduced by choosing a larger ratio $\Delta_{\max}/J_{\max}$~\cite{Privitera2018}.
In the experiment, factors such as the external confinement prevent us from choosing very slow drives and this contributes to the observation of a reduced micromotion area.

Our driving protocol including offset-sweeps can also be considered for small driving periods $T_{\rm drive}$, i.e., for near-resonant drives relative to the tunneling energy, where it does not correspond to adiabatic transport. Similar to the case with a static offset \cite{Wintersperger2020phd}, one finds a sequence of different topological phases with finite winding numbers and Chern numbers (e.g., first two spectra of Extended Data Fig.~\ref{PBC_area}d). We find that in such phases with $W_0\neq W_\pi$ the micromotion area is very small. For $T_{\rm drive}\gtrsim 6$\,ms, even though the band gaps are repeatedly opening and closing as a function of $T_{\rm drive}$, we find that the system is mostly in the same anomalous topological phase characterized by $W_0=W_\pi=2$ except for narrow regions around 'commensurate' driving periods.

\subsection{Breakdown of Thouless pump quantization}
\label{breakdown_effects}
In the 1D Thouless charge pump embedded in the hexagonal lattice, an ideal realization would lead to a quantized transport of one lattice constant per driving period. Similarly, the 2D pump protocol for the adiabatic Floquet topological phase should ideally produce a quantized value of the micromotion area. However, the experimentally observed displacement and micromotion areas are significantly reduced compared to their quantized values. We identify several effects that explain this reduction.

The first effect is the finite adiabaticity. Due to the strong external trap, we cannot perform the sequence arbitrarily slowly and this restricts the drive period range that we can choose, finding the maximal displacement for durations of the sweeps and swaps at around $\sim1$\,ms. It was shown in ref.~\cite{Privitera2018} that the suppression of band excitations for large driving periods is not exponential, but only of a power law. The exponential suppression of the Landau-Zener problem, in contrast, relies on starting far away from the avoided crossing.

The second effect comes from interactions. Within a mean-field description, the interactions modify the band structure, which can be described as a dressing of the bare bands, and which can in particular modify the band gap. Beyond-mean-field effects then lead to an additional occupation of the upper mean-field band. 
We estimate the mean-field effect in a real-space description: Interactions create an effective shift of the on-site energy given by $N_c U_H$, where $N_c$ is the number of atoms in the unit cell in the center of the sample and $U_H$ is an effective on-site Hubbard-like interaction, which we estimate to be $\sim h \cdot5.5$\,Hz. Note that interactions are repulsive and therefore more likely to hinder transport \cite{Walter2023,Pan2024}. When $N_c U_H>\Delta$ (assuming that the band gap is roughly given by $\Delta>J_{\max}$), then the energy is minimized in a configuration in which a fraction of the atoms is initially loaded into the second band (see Supplementary Information~\ref{supp::MeanField_Topology},\ref{Hubbard_U_estimation}).

The corresponding displacement can then be estimated to be scaled down by a factor $1-2f_e$, where we consider the reduced contribution from the ground state band ($1-f_e$) and the contribution with the opposite sign from the excited band $-f_e$ \cite{Aidelsburger2015,Lohse2016}.
Occupation of the second band is mainly due to interaction energy minimization when loading into the ground state of the lattice, but a finite fraction might be due to finite temperature effects \cite{Wawer2021PRA}.

Finally, while we expect the atoms to completely fill up the lowest band, we cannot exclude that a slightly inhomogeneous distribution leads to a reduced signal as most of the Berry curvature is located at the edge of the Brillouin zone \cite{Hayward2018,Lu2016PRL}. 
The expected $f_e$ differs from the datasets, e.g., of Fig.~\ref{fig:2D_Thouless}b vs. Fig.~\ref{fig:triv_2_AFTI}. This changes the scaling with respect to the $f_e=0$ case. The difference lies in the atom number $N$ and the value of $\Delta_\mathrm{max}$: $N=150.000$, $\Delta_\mathrm{max}=2\pi\cdot7.8$\,kHz (Fig.~\ref{fig:2D_Thouless}c), $N=50.000$, $\Delta_\mathrm{max}=2\pi\cdot2.6$\,kHz (Extended Data Fig.~\ref{fig:zigzag}), $N=27.000$, $\Delta_\mathrm{max}=2\pi\cdot2.6$\,kHz (Fig.~\ref{fig:triv_2_AFTI}). This corresponds to an excited fraction of 25\%, 34\%, 29\%, respectively. Inhomogeneous occupation of the band due to finite coherences can be modified by the strong interactions.

\begin{figure}[h!]
	\includegraphics[width=\linewidth]{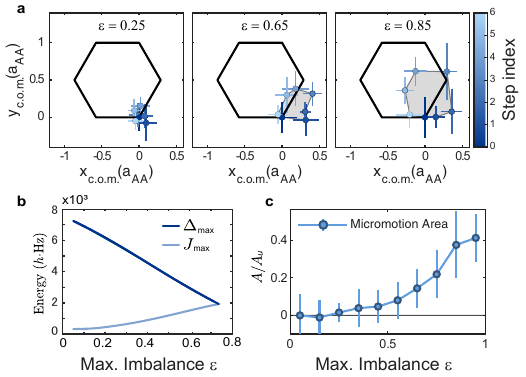}
	\caption{{\bf Influence of the tunnel modulation amplitude.} a) Center of mass position during the six steps of the protocol for the adiabatic Floquet topological system for different beam imbalances $\varepsilon=0.25, 0.65, 0.85$. b) Dependence of the maximal offset $\Delta_{\max}$ and the maximal tunneling element $J_{\max}$ on the beam imbalance $\varepsilon$. The parameters are calculated using a tight-binding description, which is valid up to $\varepsilon \lesssim  0.7$. 
    c) Measured micromotion area as a function of the beam imbalance $\varepsilon$. Errorbars indicate the $68\%$ standard deviation.}
    \label{fig:Area_vs_I}
\end{figure}

\subsection{Influence of the tunnel modulation amplitude}

We study the adiabatic realization of the anomalous Floquet topological phase for a large range of amplitudes of the tunnel modulation in order to demonstrate the robustness of the phase to parameter variation. Specifically, we fix $I_{\max}=7E_\mathrm{rec}$ and vary the beam imbalance $\varepsilon$ from $0.05$ to $0.95$. In the latter very imbalanced case, the ratio between the depths of the 1D lattices forming the total potential is given by $\sqrt{1-\varepsilon}\sim 0.2$. In the range of $\varepsilon=0.05-0.7$, $J_{\max}$ increases by a factor of 6 and $\Delta_{\max}$ decreases by a factor of 3, i.e., the ratio $J_{\max}/\Delta_{\max}$ changes by a factor of $\sim18$ (Extended Data Fig.~\ref{fig:Area_vs_I}b). 

For very small imbalances, the band gap is closed and the protocol cannot be adiabatic. For $\varepsilon\approx 0.13$, the band gap begins to open, and the micromotion area begins to rise for increasing adiabaticity (compare Fig.~\ref{fig:aAFTI_sketch}c and Section~\ref{Kitagawa_A_W}). 

For $\varepsilon>0.7$ the distance between the two minima of the lattice tends to zero during the driving period (when $I_1=I_{\max},~I_{2,3}=(1-\varepsilon) I_{\max}$ or by index permutations thereof), and we cannot reliably extract the value of the tight-binding parameters in this region. However, also in this region without two distinct potential minima, we expect anomalous Floquet topology. Because there is always a well-defined minimum in the unit cell, we would expect transport also for a classical particle following adiabatically the minimum of the potential. Still, in contrast to a sliding lattice situation, where classical particle transport is trivially realized, in this case, transport is governed by the non-trivial geometry of the time-dependent states; in fact, e.g., in the second band transport is always opposite to that of the first band \cite{Lohse2016}, as expected for a two-band system. The fact that a finite area is observed also in that region, where the strongest deformation of the potential takes place, is further evidence of the stability of the adiabatic realization of the anomalous phase.

\begin{figure}[h!]
	\includegraphics[width=\linewidth]{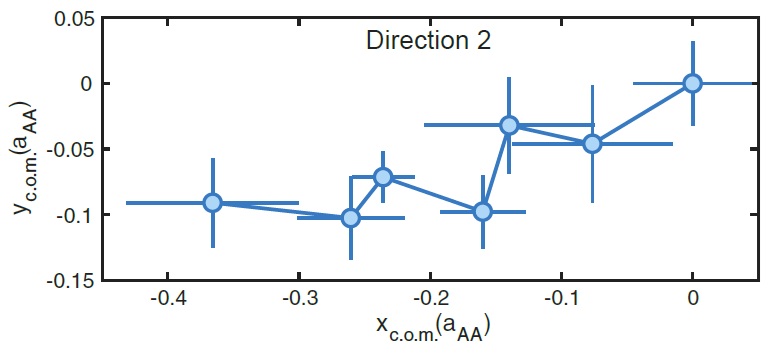}
	\caption{{\bf Motion of 1D charge pump embedded in a 2D lattice.}  Center of mass dynamics for steps $1-6$ (three periods) evidencing the zig-zag propagation as given by the underlying chain. The parameters are $I_{\max}=7\,E_r$, $\Delta/2\pi=2.6$\,kHz, $\varepsilon=0.55$, $t_{\rm sweep}=1~\mathrm{ms},~t_{\rm swap}=0.45~\mathrm{ms}$. The errorbars show the standard error of the mean.}
    \label{fig:zigzag}
\end{figure}

\subsection{Motion of 1D charge pump embedded in a 2D lattice.}
While the total displacement after three Floquet periods is well below $1~\mu$m, we are able to distinguish a zig-zag-like propagation of the center of mass in the honeycomb lattice when resolving the individual steps of the protocol (Extended Data Fig.~\ref{fig:zigzag}). The displacement is smaller than in Fig.~2, because of the smaller $\Delta_M$, where the quantized transport is affected more strongly (see Supplementary Information~\ref{supp::MeanField_Topology}). We note that the error bars showing the error of the mean calculated from 8 repetitions per point are of the same order as the zigzag signal. However, the probability that the shape of the curve of the 7 points arises by chance can be considered small.

\subsection{Matter-wave microscope} 
We start by preparing a BEC of about $1.4\cdot10^5~^{87}$Rb atoms in the $F=2,~m_F=2$ state in a magnetic trap with frequencies $\omega_{x,y,z}/2\pi=( 655,655,11)$\,Hz and adiabatically ramping up the lattice beam intensities within 100-400~ms without changing the strong $xy$ confinement of the magnetic trap. We notice that after loading into a deep 2D lattice, coherence between different lattice sites is lost (the visibility of Bragg peaks from time of flight measurements drops to zero). 

We obtain high-resolution images of the atomic density distribution in the lattice using quantum gas magnification \cite{Asteria2021}, obtained by turning off the lattice potential, letting the atomic cloud evolve for a quarter in-plane oscillation period of magnetic trap $\frac{2\pi}{4\omega_{x}} $ ($\omega_x=\omega_y=2\pi\cdot655$\,Hz), performing free-space expansion for $t_\mathrm{tof}=19.5$\,ms and finally performing absorption imaging for $50\,\mu$s. This matter-wave protocol yields a magnification factor of $\omega_x t_\mathrm{tof}=82$, for an effective resolution below $100$~nm \cite{Asteria2021}. We note that the matter-wave microscopy was also realized in the strongly-correlated regime \cite{Brandstetter2025}.

We estimate the fluctuations of the cloud center of mass due to a random position of the lattice with respect to the center of the magnetic trap from shot to shot. The associated uncertainty is about 20~nm, almost of the same order as some of the experimental error bars.

\subsection{The optical lattice}
\label{optical_lattice}
We use the tunable lattice demonstrated in \cite{Kosch2022}, where lattice beams interfere through frequency sidebands whose phases are individually controllable. It can be demonstrated that the lattice geometry depends only on one parameter, the geometry phase $\phi_g$, which is a function of the radiofrequency phases used for frequency modulation via electrooptical modulators. 
Up to translations in the $xy$ plane, the lattice potential can be written as:
\begin{equation}
    2\sum_i \sqrt{I_iI_{i+1}}\cos[({\bf k}_i-{\bf k}_{i+1})\cdot{\bf r}+\phi_g/3],
\end{equation}
where $I_i$ are proportional to the intensities of the three lattice beams in $E_r$ ($E_r=h\cdot2.02$\,kHz being the recoil energy for $\lambda=1064$\,nm and $^{87}$Rb atoms), ${\bf k}_i$ are their respective wave vectors. We identify $i=4\rightarrow i=1$ for simplicity of the notation. In the balanced lattice case $I_1=I_2=I_3$ it can be shown that $\phi_g=\pm \pi$ corresponds to a triangular lattice, and around $\phi_g\sim0$ we have a honeycomb lattice with offset $\Delta\propto\phi_g$, the proportionality constant being dependent on the lattice depth. Because $\phi_g$ depends on the radiofrequency modulation phase, we can tune the geometry on very fast time scales ($\sim\mu$s) \cite{Kosch2022}.

In this work, we do not use a balanced lattice and find that the proportionality constant between $\phi_g$ and $\Delta$ depends mostly on $\sum I_i$. So, even when the lattice balance is varied during a swap, keeping $\sum I_i$ fixed assures that also $\Delta$ does not change. Because of this, we can tune $\Delta$ and $J_i-J_{i+1}$ ($J_i$ being the tunneling couplings in the honeycomb lattice dependent on $I_i$) independently from each other, tuning $\phi_g$ for the first and the beam balance for the second. This is in contrast to Thouless pumps based on a superlattice scheme, where $\Delta$ and $J_1-J_{2}$ are a function of the relative phase of two lattices. So, the multi-frequency scheme could make it easier to realize more complex protocols with non-standard trajectories in $J_2-J_1,~\Delta$ parameter space \cite{Viebahn2024}.

\clearpage
\newpage
\section{Supplementary Information}

\setcounter{figure}{0}
\renewcommand{\figurename}{Fig. S}

\subsection{Atom number dependence of effective offset}
\label{Hubbard_U_estimation}

We estimate numerically the effective Hubbard interaction energy $U_H$  by constructing the Wannier function for the 2D lattice. In the lattice plane we approximate the wavefunction as given by the ground state of the local oscillator and in the $z$ direction we assume a Thomas-Fermi profile with a Thomas-Fermi radius calculated for about 1000 atoms residing in the same tube. The effective Hubbard interaction for two atoms that occupy such a Wannier function is calculated to be $U_H=h\cdot5.5$\,Hz. 
We directly observe the effect of such interactions in the initial state preparation by a real-space study of the density distribution. 

\begin{figure}[h!]
	\includegraphics[width=\linewidth]{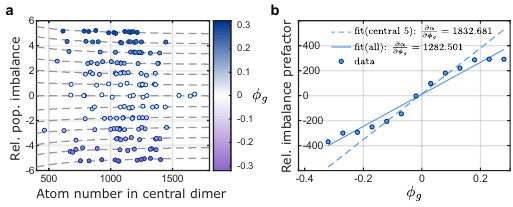}
	\caption{{\bf Relative sublattice population imbalance as a function of the geometry phase and atom number.}  a) We plot $\frac{N_a-N_b}{N_\mathrm{tot}}$ measured in the dimer in the center of the cloud. Different curves corresponds to different values of the geometry phase in the range $\phi_g/2\pi=[-0.048,0.048]$ (from top to bottom, with $\phi_g/2\pi=0.048$ corresponding to $\Delta/2\pi=\hbar\cdot7.4$\,kHz) and are offset with respect to each other for visibility. Curves are $\alpha/N$ fits to the data, where the prefactor $\alpha$ is the only fitted parameter. b) Prefactor $\alpha$ as a function of $\phi_g$. This prefactor is antiproportional to $U_H$ and proportional to the energy offset $\Delta$, which is linear in the geometry phase $\phi_g$. The expected linear dependence seems to be saturated away from $\phi_g=0$. The fitted slope to $\alpha$ considering all the points (only the central 5 points) gives a value for the effective Hubbard interaction energy $U_H$ of 15\,Hz (10\,Hz).} 
    \label{fig:AB_Imbalance}
\end{figure}

In Fig.~S\ref{fig:AB_Imbalance} we show experimental results for the population imbalance between sublattices as a function of the atom number and of the geometry phase $\phi_g$. The absolute value of the population imbalance (which changes sign when $\phi_g$ changes sign) decreases with increasing atom number. We attribute this effect to a reduced effective offset due to interactions.

1400 atoms distributed in two lattice sites with a relative energy offset of $\Delta=h\cdot2.6$\,kHz yields a relative population difference of about $\frac{\delta N}{N}=\frac{\Delta}{NU_H}=0.34$, corresponding to an occupation of the excited lattice site of $f_e=0.5-\frac{1}{2}\frac{\delta N}{N}=0.33$. We fit $\frac{\delta N}{N}=\frac{1}{N}\alpha$ as a function of $\frac{1}{N}$ for different values of $\phi_g$, where $\alpha$ is the prefactor that is antiproportional to $U_H$. We fit the dependence of the prefactor $\alpha$ with respect to $\phi_g$ and we estimate 
\begin{equation}
    U_H^\mathrm{exp}=(\frac{\partial \phi_g}{\partial \Delta}\frac{\partial\alpha}{\partial \phi_g})^{-1}=h\cdot10\,\mathrm{Hz}
\end{equation}
where $\frac{\partial \phi_g}{\partial \Delta}$ can be directly obtained from a band structure calculation and $\frac{\partial\alpha}{\partial \phi_g}$ is obtained by a linear fit in the relevant range of $\phi_g$. We attribute the discrepancy between the obtained value of $U_H$ of about $h\cdot$10\,Hz and the theoretical expectation of about $U_H=h\cdot 5.5$\,Hz to imperfections of the matter-wave protocol in the case of a honeycomb lattice \cite{Asteria2021}. In fact, we measure the populations by integrating the signal in an array of hexagonal cells of area $\frac{1}{3}A_u$ around each lattice site \cite{Kosch2022}. We detect $\sim24\%$ of the atoms landing neither in the cells corresponding to the A sites nor to the B sites (appearing as a structureless background). We expect this to arise from scattering between particles during the matter-wave optics, which is beyond a mean-field description, but assuming atoms have a probability $p_o$ of being detected outside of the lattice site where they started ($p_o$ being independent of any parameter), we can estimate the effect on the measured sublattice imbalance.
We write: 
\begin{equation}
    \begin{aligned}
\tilde{N}_a=&(1-\frac{p_o}{2})N_a+\frac{p_o}{2}N_b\\
\tilde{N}_b=&(1-\frac{p_o}{2})N_b+\frac{p_o}{2}N_a\\
\tilde{N}_c=&\frac{p_o}{2}(N_b+N_a)\\
\label{sublattices_Measured_vs_Real}
    \end{aligned}
\end{equation}
where $\tilde{N}_{a,b}$ denotes the measured atom number in the corresponding sublattice ($\tilde{N}_c$ related to the area corresponding to neither physical sublattice) and $N_{a,b}$ the "real" populations. Then subtracting the first two equations in (\ref{sublattices_Measured_vs_Real}) from each other 
\begin{equation}
    \tilde{N}_a-\tilde{N}_b=(1-p_o)(N_a-N_b)
\end{equation}
and using from the experimental observation $\frac{p_o}{2}=24\%$, we get that the observed population imbalance ($\tilde{N}_a-\tilde{N}_b$) is about half the actual one ($N_a-N_b$).
Because of this the measured imbalance fraction becomes lower and a higher than expected $U_H$ is extracted from the experimental measurement, and the obtained factor of 2 seems to explain quite well the deviation between theory ($5.5$\,Hz) and experiment ($10$\,Hz in the experimentally relevant region of $\phi_g$).
We do not expect interactions to lead to population of even higher bands as they are separated by a band gap of $\sim14$\,kHz.

\subsection{Measurement of adiabaticity}
\label{supp::adiabaticity_measurement}

\begin{figure}[h!]
	\includegraphics[width=\linewidth]{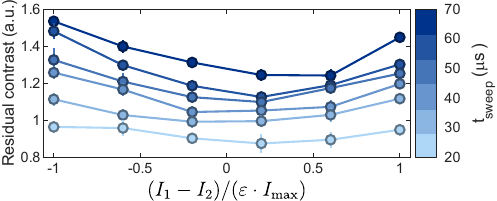}
	\caption{{\bf Lattice contrast after Landau Zener sweep.}
    {Lattice contrast measured in real space as a function of $\frac{I_1-I_2}{\varepsilon I_\mathrm{max}}$ during a fast sweep of $\Delta$ from very negative to very positive values, for different sweep times. The lattice contrast is lowest around $I_1-I_2\sim0$ indicating more transfer of atoms to higher bands which are not sharply resolved in real space.}}
   \label{fig:lattice_contrast}
\end{figure}

We characterize the instantaneous band gap by measuring the adiabaticity of different parameter ramps.  
We start by adiabatically loading into a triangular lattice, which is characterized by $\phi_g=-\pi$ \cite{Kosch2022}, for fixed $I_3=(1-\varepsilon) I_{\max}$ and $I_1+I_2+I_3=(3-2\varepsilon)I_{\max}$ and for a variable value of $I_1-I_2$. Secondly, we perform a ramp of $\phi_g$ to $\phi_g=\pi$, for variable sweep times $t_\text{sweep}$. We then ramp up the lattice intensities to $I_i=I_{\max}$, allow some time for thermalization and measure the density distribution in situ. Fast ramps ($\sim10-20\,\mu$s) result in high excitation probability to higher bands \cite{Wirth2011,Kosch2022}, and this is seen in real space by measuring the lattice contrast. This is defined as the mean quadratic deviation between the cloud profile integrated along $y$ (and naturally along $z$ due to the nature of the imaging), and a gauss fit to the profile itself. The deviations are bigger for clouds where more atoms are in the lowest bands because they exhibit the biggest lattice contrast \cite{Asteria2021}. So the lattice contrast is used as an indirect measure of adiabaticity, because faster sweeps or sweeps in presence of a smaller band gap lead to higher bands population and hence to a smaller lattice contrast. For longer sweep times, a minimum of the lattice contrast emerges when $I_1-I_2\sim0$ (Fig.~S\ref{fig:lattice_contrast}), consistent with the fact that there the instantaneous band gap closes in the middle of the sweep for $\phi_g=0$ (i.e. $\Delta=0$). Note that the ratio between sweep amplitude and sweep time used here is 2-3 orders of magnitude bigger than the one used in the main experiment, where we do not observe any detectable loss of contrast as a function of time during one Floquet period. 

\subsection{Numerical simulations}
\label{supp::numerics}
\subsubsection{Simulation of the experimental protocol}
The enclosed micromotion area $A$ and the quantity $d^2$ as obtained numerically and reported in Fig.~3 are computed as follows. We simulate in real space the micromotion dynamics starting from a single lattice site. We consider the tight binding parameters $\Delta_\mathrm{max}=2\pi\cdot2.6$\,kHz, $J_\mathrm{max}=1.03$\,kHz. Because of the non-linear relation between $I_i$ and $J_i$ during a linear ramp  of the beam intensities the tunneling couplings vary (slightly) non-linearly in time. We observe that there is basically no dependence on the amount of non-linearity and for simplicity we consider a linear ramp of the tunneling couplings as well. Because at $t=0$ interactions compensate the potential shift due to the harmonic confinement (to minimize the total energy of the system) in a first approximation the harmonic confinement can be neglected and we don't consider it here. Also, 
as elaborated in Section~\ref{supp::MeanField_Topology}, as the residual interaction effects are relatively small, we do not consider interactions either. From these simulations, we directly extract $A$ and $d^2$ as a function of the path width. The micromotion area is defined such that the final and initial points of a period are connected by a straight line to ensure a closed area. The results presented in Fig.~3 are scaled down by a factor of 0.33 to take into account the effects described in Section~\ref{supp::MeanField_Topology}.

\subsection{Definition of the winding number in adiabatic protocols}
\label{adiabatic_Winding_Number}
The winding number $W_g$, which counts the number of edge modes in a gap $g$ is defined as \cite{Rudner2013}:
\begin{equation}
\begin{aligned}
    W_{g}=&\frac{1}{8\pi^2}\int_0^{2T} dt \int_{BZ}\mathrm{d}{\mathbf{k}}~\mathrm{Tr}~~\hat{O}\\
    \hat{O}=&-{U_g}^{-1}\partial_t {U_g}~ \Bigl[~{U_g}^{-1}\partial_x {U_g}, {U_g}^{-1}\partial_y {U_g}~\Bigr]
    \end{aligned}
\end{equation}
with ${U_g}$ being the band-flattened time-evolution operator (in this case, a $n_b$ by $n_b$ matrix, where $n_b$ is the number of bands involved). ${U_g}$ depends on the wavevector (in the Brillouin zone, denoted by $BZ$) and the evolution time $0<t<2T$ where $T$ is the Floquet period. It is defined as \cite{Rudner2013}:
\begin{equation}
 {U}_g(\bk,t)=
\begin{cases}
U(\bk,t),&0\le t\le T\\
\exp\big(-\frac{i}{\hbar}(2T-t)H_f^g(\bk)\big), & T<t\le 2T,
\end{cases}
 \end{equation}
 where
\begin{equation}
  {{H}^g_f(k)=\frac{i\hbar}{T}\log\bigl[{U}(\bk,t=T)~\bigr]_\text{branch cut at $g/T$}.}
\end{equation} 
${H}_f^g$ is the Floquet Hamiltonian, calculated from the physical time-evolution operator $U$. The Floquet Hamiltonian depends on the band gap $g$: the branch cut of the matrix logarithm can be put in each of the quasienergy bandgaps $g$. Correspondingly, $W_g$ obtained by placing the branch cut in the $g$ gap will count the number of edge modes residing into the $g$ gap. Note that the number of gaps in a driven system is given by $n_b$, and there are correspondingly $n_b$ independent $W_g$.
Note also that ${U}$ is dependent not only on $t$ and $k$, but also on $g$, and for readability, we omit these dependencies.
For $t<T$, ${U}$ corresponds to the time-evolution operator which describes the system's dynamics during the Floquet drive.
For $t>T$, an artificial dynamics is introduced, driven by the Floquet Hamiltonian, which does not change the number of edge modes in the gap $g$ but is used to ensure that ${U}(t=2T)={U}(t=0)=\mathbb{I}$. Thus periodic boundary conditions are obtained also in the time dimension and $W_g$ is well-defined \cite{Rudner2013}. 

We assume non-degenerate bands at any point in time. In the infinitely slow parameter change associated to an adiabatic protocol, we can replace ${U}$ with $\tilde{U}$ which is defined as 
$\tilde{U}=\sum_je^{i\theta_j(T)}\ket{\psi_j(t)}\bra{\psi_j(0)}$. 
$\theta_j$ is the phase picked up during evolution (consisting of a part dependent on the integral of the energy over time and a time-independent geometrical part), and $\psi_j(t)$ are the time- and quasimomentum dependent eigenstates of the instantaneous Hamiltonian $H(t)$.
At $t=T$, $\tilde{U}=\sum_j e^{-i\theta_j(T)}\ket{\psi_j(T)}\bra{\psi_j(0)}=\sum_j e^{-i\theta_j(T)}\ket{\psi_j(0)}\bra{\psi_j(0)}$.
It can be directly seen, that $H_f^g$ has the same eigenstates as $H(0)$, the instantanous Hamiltonian at $t=0$. Assuming that the bands of $H(0)$ have all Chern number 0, it follows that also the bands of $H_f^g$ have Chern number 0. As a consequence, $W_g=W_{g'}$ for every $g,g'$, because $W_{g+1}-W_g$ is given by the Chern number of the band residing between the  $g$ and $g+1$ gap.

When modifying $\tilde{U}$ by choosing a slightly different path in parameter space, the eigenstates of $H_f^g$ given by $\theta_j$ might exhibit some accidental degeneracy. Such band-gap-closing points {in the Floquet Hamiltonian} do not correspond to a topological phase transition, as the relevant winding number associated to the closing band gap $W_g$ cannot change as it is forced to assume the same value of all the other $W_{g'}$. 
In conclusion, the path in parameter space can be adiabatically deformed without changing the value of $W_g=W_{g'}$. 

To calculate the winding number for a given path, it is convenient to deform the path to the associated point of  dispersionless micromotion dynamics (\cite{asteria2026micromotionareaproxyanomalous}), where the winding number can be obtained by a simple calculation or visualized directly \cite{Kitagawa2010,Rudner2013}.

In particular, in the case $n_b=2$, it was shown \cite{asteria2026micromotionareaproxyanomalous} that the winding number is given at the dispersionless point by $2\frac{A}{A_u}$ where $A$ is the area enclosed by the cyclotron orbit in the bulk and $A_u$ is the unit cell area of the lattice. In the experiments presented here, since the enclosed area during one Floquet period at the associated dispersionless point is $A=A_u$, we expect $W_0=W_\pi=2$.

\subsection{Topology with mean-field interactions and a harmonic trap}
\label{supp::MeanField_Topology}

The winding number is defined on the time evolution operator in momentum space for a non-interacting system with periodic boundary conditions. In our experimental case, in turn, we have a strong harmonic confinement and non-negligible interactions. 

We consider an incoherent superposition of the two lowest bands of the lattice, and within each band a uniform occupation of all the momenta.
This is motivated by the presence of interactions and experimentally confirmed by the absence of visibility of Bragg peaks in time-of-flight pictures.

We define $f_e$ as the excited fraction in the second band. We consider now a translationally invariant system and a variable atom number per unit cell $N_c$ up to $\sim2000$. The global state at $t=0$ depends on $f_e$ and $N_c$. In particular we are interested in determining the sublattice offset correction due to interactions $\tilde{\Delta}=\delta N U_H$. $U_H$ is an effective Hubbard on-site interaction energy \cite{zahn2021formation}
and we estimate $U_H=h\cdot5.5$Hz in Section~\ref{Hubbard_U_estimation} for the $7E_r$ deep lattice. $\delta N$ is the population imbalance between sublattices ranging from $-N_c$ to $N_c$ and is defined as 
\begin{equation}
    \delta N(t)=N_c(1-2f_e)~\Bigl[~1-2\overline{\sin(\theta(\mathbf{k},t)/2)^2}~]
\end{equation}
where $\theta=[0,\pi]$ is the polar angle in the Bloch sphere representation. The average is taken over the Brillouin zone. $\delta N$ and $\theta$ are also functions of time. 
{For $t=0$, we assume $\delta N(t=0)\sim N_c(1-2f_e)$, because in the limit $\Delta_{\max}\gg J_{\max}$ we have $\sin(\theta(\mathbf{k},t=0))\sim0$.}

At $t\neq0$, we consider the time evolution (which determines $\theta(\mathbf{k},t),~\phi(\mathbf{k},t)$) given by the time-dependent single particle Hamiltonian as described by $\Delta(t),~J(t)$ by replacing $\Delta$ with $\Delta+\tilde{\Delta}$ to take interactions into account. $\delta N$ can be rewritten as:
 \begin{equation}
     \delta N(t)=\frac{\delta N(t=0)}{1-2\overline{\cos(\theta(\mathbf{k},t=0)/2)^2}}~\Bigl[~1-2\overline{\cos(\theta(\mathbf{k},t)/2)^2}~\Bigr]
\end{equation}
which no explicit dependence on $N_c$. Therefore the time evolved states will only be a function of $\delta N$ and not $N_c$. 

In the experiment with a tightly confining trap, $\delta N$ between nearest neighbors is determined by $\Delta$ and $U_H$ and pretty much independent of the position within the cloud. Therefore, we can approximate the dynamics as being described by the same $\delta N(t)$ everywhere even if $N_c$ is position dependent. The only role played by $N_c$ is in the determination of $f_e$, which is then averaged over the system in order to get the correction between displacement associated to a singly occupied band and the observed displacement, due to the fact that the two bands are characterized by opposite velocities (and hence the correction factor is given by $1-2f_e$). 

For the dataset of figure 2b, $f_e$ is estimated to be quite close to 0.5 which qualitatively explains why the observed displacement is much smaller than the expected one for a non-interacting system. 
In contrast to the data presented in figure 3, for figure 4 of the main text, $f_e$ is not constant as it depends on $I$. In particular, as $\Delta_{\max}$ depends on $I$, the initial fraction of atoms in the second band depends on $I$ and this effect is taken into account when calculating the effective interactions and for rescaling the observed real space dynamics. 
We calculate the excited fraction in the initial state in the center of the trap according to 
\begin{equation}
    f_e^{\mathrm{0}}=\mathrm{max}\Bigl[0,\frac{1}{2}(1-\frac{\Delta_{\max}}{N_cU_H})\Bigr]
\end{equation}

The excited fraction in the total system is then estimated as follows.  We use a Thomas Fermi distribution assumption:
We approximate the number of atoms in the lowest band $N_a$ as a homogeneous continuous distribution and rotationally symmetric as given by the integral:
\begin{equation}
    N_a=\alpha\int_0^1dr~r(1-r^2)
\end{equation}
where $r$ is the radius in units of the Thomas-Fermi radius. The factor $1-r^2$ accounts for the Thomas-Fermi distribution and the factor $r$ is obtained after integration over the in-plane angle.
The number of atoms in the second band is given by: 
\begin{equation}
    N_b=\alpha\int_0^{\sqrt{p_0}}dr~r(p_0-r^2)
\end{equation}
where $p_0$ is the ratio between second and first band in the center of the trap. 
Solving the integrals we obtain
\begin{equation}
    p=\frac{N_b}{N_a}=\frac{\alpha(\frac{p_0^2}{2}-\frac{p_0^4}{4})}{\frac{1}{4}\alpha}=2\frac{p_0^2}{2}-{p_0^4}
\end{equation}

Where $p$ is the ratio of atoms in the second  band to atoms in the first band.
In the end we use:
\begin{equation}
\begin{aligned}
f_e=\frac{1}{1+p}
\end{aligned}
\end{equation}
to get the excited state fraction $f_e$ in the whole system.

Displacements of the center of mass are rescaled by a factor $1-2f_e$ (considering the $1-f_e$ contribution from the first band and the $-f_e$ contribution with the opposite sign for the excited band. The minus sign comes from the fact that the velocity and position operators have a Berry-curvature-like expression, and the Berry curvature is opposite for the two bands \cite{asteria2026micromotionareaproxyanomalous}) and areas are rescaled by a factor $(1-2f_e)^2$. 
We consider a reduction factor of 0.7 in the effective $U_H$ during the dynamics to account for the fact that the Thomas-Fermi radius in $z$ direction is quite different in the two sublattices, and the two corresponding density distributions have a reduced overlap estimated to be around 0.7. This is motivated by the fact that the dynamics in $z$ direction is slower than a driving period.

\begin{figure}[h]
\includegraphics[width=\linewidth]{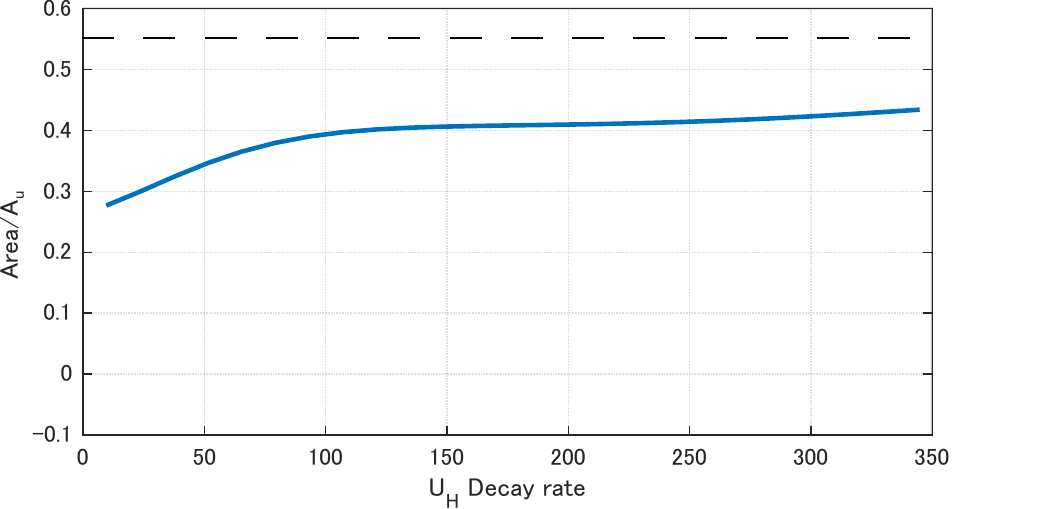}
\caption{{\bf Area as a function of $\Gamma_U$}. Area (blue curve) as a function of the decay rate $\Gamma_U$ of the effective on-site interaction strength $U_H$. Parameters correspond to Fig.~3 for the maximal path width. For $U_H=0$, equivalent to $\Gamma_U=\infty$, we get $A/A_u=0.55$ (dashed line) \label{U_decay}}
\end{figure}

We also consider a time-dependence of $U_H$ when initiating the dynamics. This is motivated by the fact that the atom number in each tube is changed quite fast during the sweeps as compared to the timescale in the tube direction given by $\omega_z=2\pi\cdot30$Hz and therefore adiabaticity is not maintained with regard to this degree of freedom. Expansion in the tube direction driven by interactions leads to a decrease of $U_H$ as a function of time. We model this decrease heuristically with a negative exponential of rate $\Gamma_U$. We cannot access with high precision the momentum distribution along $z$ direction, and have therefore no experimental indication about which value of $\Gamma_U$ best describes our experiment. A naive expectation would be $\Gamma_U\sim\frac{2\omega_z}{\pi}\sim133$Hz as the inverse of the $T/4$ time in the $z$ direction. Numerical simulations show that already at $\Gamma_U\sim100$Hz the area has reached 70\% of the non-interacting value (which corresponds to $\Gamma_U=\infty$). We plot in Fig.~S\ref{U_decay} the value of the area for parameters corresponding to Fig.~\ref{fig:triv_2_AFTI} for the maximal path width.

From this analysis, we can demonstrate the robustness of the adiabatic realization against interactions. Considering as an example the case $\Gamma_U\sim100$Hz, the average interaction energy associated to $\Delta N$ is $\Delta NU_H(t)$. Averaged over time, the interaction energy is still $\sim1000\cdot2\pi\cdot0.7\cdot5.5\mathrm{Hz}\cdot(1-e^{-1})\sim2\pi~2.5\mathrm{kHz}$, which is 2.5 times the kinetic energy scale given by $J_{\max}=2\pi\cdot1\mathrm{kHz}$ and directly comparable with the sublattice offset $\Delta_{\max}=2\pi\cdot2.6\mathrm{kHz}$.

There might be a few other effects which may be responsible for deviations between simulations and experiment. We consider a fixed Thomas-Fermi radius in $z$ direction which is position independent (and consequently $U_H$). The local density approximation might not be accurate when atoms travel even for a few lattice vectors in such a tight trap. In general, we are aware that the description of this unusual system geometry, with finite temperature, interactions and time-dependent potentials can be quite challenging for mean-field theories \cite{zahn2021formation}, and there might be effects that are not captured by such numerical methods \cite{Chen2011,Jalali-Mola2026}.
{Similarly to the discussion for quantized transport of bosons via Thouless pumps, a non-uniform occupation of the Brillouin zone in the case of bosonic atoms might also play a role.}

\subsection{Quantization of transport in adiabatic protocols in the 2D case}
\label{supp:Chern_T}
In infinite 1D systems, as proven by Thouless \cite{thouless1983}, adiabatic particle transport is quantized in terms of the Chern number $C_T$ defined as 
\begin{equation}
\begin{aligned}
    C_T=\frac{i}{2\pi}\int_0^T\mathrm{d}t\int_{-\pi}^\pi dk_x~\Bigl[~\bra{\frac{\partial \psi(t,k_x)}{\partial k_x}}\ket{\frac{\partial \psi(t,k_x)}{\partial t}}&\\-\bra{\frac{\partial \psi(t,k_x)}{\partial t}}\ket{\frac{\partial \psi(t,k_x)}{\partial k_x}}&\Bigr]
    \end{aligned}\label{Chern_Thouless}
\end{equation}
where $\psi(t,k_x)$ indicates the eigenstates of a uniformly filled band as a function of quasimomentum $k_x$ and time $t$ (where we omitted for simplicity the sum over all filled bands). If adiabaticity is achieved, the wavefunction follows the instantaneous eigenstate and the Chern number of the protocol dictates the amount of pumped charge along the direction $x$. The expression for the Chern number given here is identical to the one used for the Chern number of 2D systems $C_{2D}$, with the substitution of the time dimension for a second spatial dimension. In that case, the Chern number $C_{2D}$ counts the difference between the number of edge modes above and below a certain band. 
We consider a 2D system which is periodically driven slowly compared to its energy scales, which is to say the instantaneous wavefunction follows adiabatically a given eigenstate (for simplicity, the ground state), identified by $\psi(t,k_x,k_y)$. This assumption of adiabaticity can be fulfilled as long as the interchain tunneling does not lead to a closing of the band gaps. We consider the displacement of a filled quasimomentum "strip" at a fixed value of $k_y$, $\delta x(k_y)$:
\begin{equation}
    \begin{aligned}\delta x(k_y)=\frac{i}{2\pi}\int_0^T\mathrm{d}t\int_{-\pi}^\pi dk_x~&\Bigl[~\bra{\frac{\partial \psi(t,k_x,k_y)}{\partial k_x}}\ket{\frac{\partial \psi(t,k_x,k_y)}{\partial t}}\\-&\bra{\frac{\partial \psi(t,k_x,k_y)}{\partial t}}\ket{\frac{\partial \psi(t,k_x,k_y)}{\partial k_x}}~\Bigr].
    \label{delta_x_2D}
    \end{aligned}
\end{equation}
Because Eq.~(\ref{delta_x_2D}) is equivalent to Eq.~(\ref{Chern_Thouless}), $\delta x(k_y)$ must be quantized. Since $\delta x(k_y)$ is a smooth function of $k_y$, and adiabaticity is assumed for every value of $k_y$, then it follows that $\delta x$ does not depend on $k_y$. The total displacement along $x$, calculated by averaging over $k_y$ must therefore also be quantized. We attribute the deviations from the quantized value to the issues discussed in the main text (interactions, strong confinement).

\subsection{Numerical implementation of the Floquet-Simulation}
 
The experimental protocol studied in Section~\ref{experiment_adafti} of the main text essentially consists of 6 Landau-Zener sweeps between the A- and B-sublattices interspersed by a cyclic swapping of the dominant tunneling direction. In the Methods~\ref{Kitagawa_A_W}, we numerically study a variation of this protocol as a function of the driving time $T_{\rm drive}$. In this simulation, we set the swap time to zero and consider a protocol consisting of sweeps followed by instantaneous cyclic swapping of the active tunneling direction, where the tunneling in the other two directions is set to zero. The total period of the protocol is then $T_{\rm drive} = 6\cdot t_{\rm sweep}$. 

In the numerics, we calculate the Floquet operator to get the quasi-energy spectrum. During each step the instantaneous two-band Hamiltonian for a given quasi momentum $\vec{k}$ is given by
\begin{equation}
    \mathcal{H}_{AB}^\alpha(\vec{k}, \tau) = \begin{pmatrix}\Delta(\tau)/2 & -J_{\mu_\alpha} e^{-i\Phi_{\mu_\alpha}(\vec{k})} \\ -J_{\mu_\alpha} e^{i\Phi_{\mu_\alpha}(\vec{k})} & \Delta(\tau)/2\end{pmatrix}
\end{equation}
with the active tunneling direction $\mu_\alpha = (2,3,1,2,3,1)$ for $\alpha= 0,...,5$ and $\Phi_{\mu_\alpha} = \vec{k}\cdot\vec{\delta}_{\mu_\alpha}$ the k-dependent phase for tunneling in direction $\vec{\delta}_{\mu_\alpha}$, which are the three vectors connecting one A site with the three neighboring B sites. We define the parameter $\tau_\alpha = t - (\alpha +\frac{1}{2} ) t_{\rm sweep}$ as the time parameterizing each sweep, which runs from $-0.5\, t_{sweep}$ to $+0.5\, t_{sweep}$ within each sweep. The time-dependent sublattice offset is then given by $\Delta(\tau_\alpha) = (-1)^\alpha \beta \tau_\alpha $ with the sweep velocity $\beta = 2\Delta_{\rm max}/t_{\rm sweep}$. The Floquet operator can by expressed explicitly as
\begin{equation}\label{Ufloquet}
    \mathcal{U}_F(\vec{k}) = \mathcal{T}_\alpha \prod_\alpha R(\Phi_{\mu_\alpha}(\vec{k})) \,u_\alpha(t_{\rm sweep}) \, R(\Phi_{\mu_\alpha}(\vec{k}))^\dagger
\end{equation}where $u(t_{\mathrm{\rm sweep}})$ denotes the propagator for the finite time Landau-Zener sweep in step $\alpha$, $\mathcal{T}_\alpha$ denotes the time-ordering w.r.t. the steps and conjugation with the diagonal matrices
\begin{equation}
    R(\Phi_{\mu_\alpha}(\vec{k}))=\begin{pmatrix}e^{-i\Phi_{\mu_\alpha}(\vec{k})/2} & 0 \\ 0 & e^{i\Phi_{\mu_\alpha}(\vec{k})/2}\end{pmatrix}
\end{equation}
adds the $\vec{k}$-dependent phase for tunneling along the direction vector $\vec{\delta}_{\mu_\alpha}$
\subsubsection{Derivation of the LZ-Propagator}
Expressing the time dependent Schrödinger equation for the two level system with an energy offset $\Delta$, which is varying linearly in time with a rate $\beta$ in the diabatic basis $\{|A\rangle, | B \rangle\} $, one can transform the two first order differential equations for the coefficients
\begin{align}
        i \dot c_A &= \frac{\beta \tau}{2}c_A -J c_B \\ \label{cBdot} i \dot c_B &= -J c_A -\frac{\beta \tau}{2}c_B
\end{align}
into a second order equation~\cite{shevchenko_landauzenerstuckelberg_2010}
\begin{equation}
    \ddot c_A = (-i \frac{\beta}{2} -J^2 - \frac{\beta^2 \tau^2}{4}) c_A
\end{equation}
Introducing the variable $z\coloneq  e^{-i\frac{\pi}{4}}\sqrt{\beta}\tau$ and $\gamma \coloneq \frac{J^2}{\beta}$ this equation can be brought into the form of the Weber equation 
\begin{equation}\label{eq:weber_eqn}
    \dv[2]{c_A}{z} + (\nu_A + \frac{1}{2} -\frac{z^2}{4})c_A(z) = 0,
\end{equation}
with $\nu_A = 1-i\gamma$, with solutions given by the parabolic cylinder functions (PCF) $D_{\nu_A}(s z)$, with $s=\pm1$. For the second component $c_B$ we can obtain a solution by using Eq.~(\ref{cBdot}) and an identity for the derivative for the PCFs
\begin{equation}
    \dv{\lambda} D_\nu(\lambda) = \frac{\lambda}{2} D_\nu(\lambda) - D_{\nu +1}(\lambda) 
\end{equation}
We end up with 
\begin{equation}
    \begin{aligned}
        c_B &= \frac{1}{J}(\frac{\beta \tau}{2} c_A - i \dot c_A)\\
            &= \frac{1}{J}(\sqrt{\beta}\,e^{i\pi/4}\, \frac{z}{2} D_{\nu_A}(s z)\\
            &-i\sqrt{\beta}\,e^{-i\pi/4}\,(\frac{z}{2}D_{\nu_A}(s z)- s D_{\nu_A+1}(s z))\\
            &= \frac{e^{-i3\pi/4}}{\sqrt{\gamma}} s D_{\nu_B}(s z(\tau))=\zeta  s  D_{\nu_B}(s z(\tau)) 
    \end{aligned}
\end{equation}
where we have defined $\zeta = (\sqrt{\beta}e^{i3\pi/4})^{-1}$ and used $ie^{-i\pi/4} = e^{i\pi/4}$, canceling the first two terms and leaving us with the solution. With these we can construct the fundamental Matrix $M(\tau)$ with columns given by the two independent solutions to Eq.~(\ref{eq:weber_eqn})
\begin{equation}
M(\tau) =  \begin{pmatrix} D_{\nu_A}(z(\tau)) &D_{\nu_A}(-z(\tau)) ) \\ \zeta \,D_{\nu_B}(z(\tau)) &  - \zeta \,D_{\nu_B}( -z(\tau))\end{pmatrix}
\end{equation}
which defines gives the propagator from time $\tau_i$ to $\tau_f$
\begin{equation}
    u(\tau_f, \tau_i) = M(\tau_f)M(\tau_i)^{-1}
\end{equation}
With this we compute the Floquet operator by evaluating the factors in Eq.~(\ref{Ufloquet}) for each step  using the \textit{mpmath} python library for arbitrary precision floating point arithmetic \cite{mpmath}.

\end{document}